\documentclass[journal,comsoc]{IEEEtran}
\usepackage{amsmath,amsfonts}
\usepackage{algorithmic}
\usepackage{algorithm}
\usepackage{array}
\usepackage[caption=false,font=normalsize,labelfont=sf,textfont=sf]{subfig}
\usepackage{textcomp}
\usepackage{stfloats}
\usepackage{url}
\usepackage{verbatim}
\usepackage{color}
\usepackage{graphicx}
\graphicspath{{Schematics/}} 
\usepackage{cite}
\begin{document}

\title{Data-oriented Coordinated Uplink Transmission for Massive IoT System}

\author{Jyri H\"am\"al\"ainen,~\IEEEmembership{Senior Member,~IEEE,} Rui Dinis,~\IEEEmembership{Senior Member,~IEEE}, Mehmet C. Ilter,~\IEEEmembership{Senior Member,~IEEE}}
\maketitle


\maketitle

\begin{abstract}
Recently, the paradigm of massive ultra-reliable low-latency IoT communications (URLLC-IoT) has gained growing interest. Reliable delay-critical uplink transmission in IoT is a challenging task since low-complex devices typically do not support multiple antennas or demanding signal processing tasks. However, in many IoT services the data volumes are small and deployments may include massive number of devices. We consider on a clustered uplink transmission with two cooperation approaches: First, we focus on scenario where location-based channel knowledge map (CKM) is applied to enable cooperation. Second, we consider a scenario where scarce channel side-information is applied in transmission. In both scenarios we also model and analyse the impact of erroneous information. In the performance evaluation we apply the recently introduced data-oriented approach that has gathered significant attention in the context of short-packet transmissions. Specifically, it introduces a transient performance metric for small data transmissions, where the amount of data and available bandwidth play crucial roles. Results show that cooperation between clustered IoT devices may provide notable benefits in terms of increased range. It is noticed that the performance is heavily depending on the strength of the static channel component in the CKM based cooperation. The channel side-information based cooperation is robust against changes in the radio environment but sensitive to possible errors in the channel side-information. Even with large IoT device clusters, side-information errors may set a limit for the use of services assuming high-reliability and low-latency. Analytic results are verified against simulations, showing only minor differences at low probability levels. 
\end{abstract}


\section{Introduction}
\IEEEPARstart{T}{he} Internet of Things (IoT) is an umbrella term for concepts connecting various physical objects and devices to the internet and allowing them to collect and exchange data. While IoT is an critically important field for future research, several comprehensive surveys with different focus areas and aspects of IoT have been published including 5G IoT \cite{ChBe2020} and 6G IoT \cite{NgDi2022}, as well as many special topics such as ultra low-power communications \cite{Jiang2023}, machine-type communications \cite{ShWa2020}, IoT system autonomy \cite{AsTa2023}, to name few. 

As noticed in the cellular communication research community, the IoT equipment can be divided into 'high-end' and 'low-end' devices with different kind of quality of service (QoS) requirements. The wireless collection of environmental data, for example, is typically carried out by sensors that on one hand admit low transmission power on narrow bandwidth but, on other hand, the collected data is not aging in milliseconds and communication is delay tolerant. In contrary, if, e.g., sensor data from a large machinery is used for wireless control purposes, then demanding QoS requirements in terms of reception reliability and low communication latency are faced. While delay tolerant IoT communications has been extensively studied, the paradigm of massive {\it ultra-reliable low-latency IoT communications} (URLLC-IoT) has also gained growing interest recently. In 5G domain the discussion on URLLC already covered some IoT use cases \cite{Nas2019} but topic is more integral part of the 6G development \cite{NgDi2022},\cite{AdBe2020}.   

Executing a reliable delay-critical transmission in IoT is a challenging task. The low-complex devices are typically not supporting multiple antennas and demanding signal processing tasks. Yet, there are two aspects that can help supporting URLLC in the IoT domain. First, in many IoT services the data volumes are small. Second, IoT deployments may include massive number of devices. Due to small data volumes the maximisation of link capacity is not as important as in e.g. massive MIMO applications but emphasis is merely in the link reliability. Furthermore, the large number of IoT devices potentially support cooperative transmission approaches even though it must be taken into account that cooperation cannot assume very sophisticated methods or radio resource intensive information exchange between receiver and transmitting devices. Accordingly, we focus on an approach where a reliable low-latency uplink transmission is executed by a group of clustered and cooperating IoT devices. We expect that successful cooperative transmission requires only scarce location or channel side-information. Moreover, an important aspect of our study is that we assume incomplete and erroneous side-information in the transmitting cluster. 

\subsubsection{Channel knowledge map based approach} While joint transmission is typically based on either direction information or channel feedback from the receiver, recently a new concept of CKM has been introduced \cite{ZeXu2021}. The CKM represents a development of the radio and/or built environment aware communications, see e.g. \cite{EsGa2019,BiLy2019}, providing from a database certain channel-related information for transmitter-receiver pairs. It is envisioned that CKM-enabled communications could even replace the need for real-time channel state information (CSI) when e.g. channel path information for a specific transmitter-receiver pair is known, including the number of significant channel paths and their powers, phases and delays \cite{WuZe2021}. From CKM perspective a favourable scenario takes place when there is a strong LoS between transmitter and receiver. In such a situation the path loss between transmitter and receiver can be easily computed if locations are known. However, since IoT devices are typically power limited, the path loss information as such may provide only understanding on the power gap that transmitter needs to overcome for a successful reception. To tackle this kind of challenge cooperative transmission methods can be used especially if IoT devices occur in clusters.              

\subsubsection{Feedback based cooperation} Since CKM based approach has its limitations, we will also consider a more conventional feedback based cooperation in cluster of IoT devices. Even though multiple-input single-output (MISO) with co-located antennas is a well studied topic \cite{Mol2011}, \cite{GoAs2012}, the use of explicitly defined simple methods with quantised channel information has not been extensively studied. Such methods are typically suboptimal but they provide a good insight on the impact of feedback quantisation and erroneous channel information \cite{Alexis2015,Alexis2016,Ben2015,Ham2002}. Low resolution feedback is an especially attractive choice for IoT communications because therein a limited reverse channel capacity may set strict constraints for feedback data volume. When compared to CKM-based approach the use of explicit feedback from receiver has a clear advantage if LoS is weak and channel contains a fast fading component. On the other hand, the drawback is that obtaining channel information in the receiver may require training or sounding signals from IoT devices consuming both radio and power resources of IoT devices. Finally, feedback information is vulnerable to errors in the reverse channel and, in addition to power consumption, the channel sounding signal from an individual IoT device might be too weak for a reliable reception in the receiver.         

\subsubsection{Delay Outage Rate}

To assess the proposed system feasible key performance indicators (KPIs) are needed. In this respect, a new concept called as data-oriented approach was introduced for the design and analysis of advanced wireless systems \cite{YaAl2019}, \cite{YaCh2020} where data transmission sessions complete within
one channel coherence time over short packet transmission
sessions. Basically, this concept is simply representing a new way to look at the link performance from the transmission delay outage point of view and authors in \cite{YaAl2019} named it as the {\it Delay Outage Rate} (DOR) which refers the probability that the required information delivery time for a specific transmission exceeds the predefined threshold. The DOR metric for the short-packet transmission over fading channels was presented in \cite{Yang2020} and inspired by it, \cite{Ilter2021} showed how the DOR expression can be utilized for searching optimum constellation points over a coded scenario. 
Furthermore, the DOR metric has been recently utilized in the coverage analysis of reflective intelligent surfaces (RIS)-aided communication systems \cite{Yang2020RIS} and in visible light communication scenarios \cite{Singh2022}. In \cite{can2023}, the data-oriented approach was adopted into downlink RSMA systems where two different precoder designs were investigated based on the proposed DOR metric.

\begin{figure}[H]
\centering
\includegraphics[width=9cm]{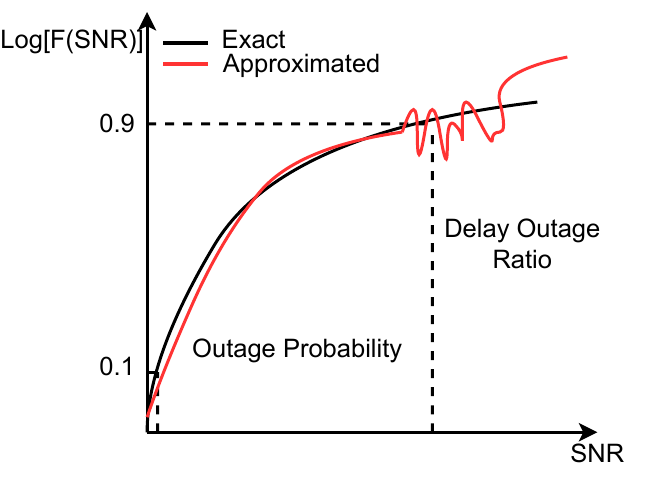}
\caption{The illustration of DOR and outage probability.}
\label{fig:DORvsOutage}
\end{figure}

The mentioned DOR studies provide a new perspective on the wireless link performance and it can be used to evaluate simultaneously reliability and latency. In simplest setting the DOR study can be carried out straightforwardly by using the cumulative distribution function (CDF) of the signal-to-noise-ratio (SNR). In all mentioned studies, obtaining the DOR expression basically falls back to finding the CDF of received SNR. While, in this respect, it can be considered as a simple extension of the outage probability analysis, there is a notable technical difference. Namely, study of high reliability and low latency require accurate knowledge of the tail of the CDF of SNR. If the exact distribution is known, then this is not a problem but in case when the CDF is obtained as an approximation it is typical that approximation works well in 0.1 - 0.9 probability levels but when looking at the tail's fit in the logarithmic scale notable errors may occur as illustrated in Fig.\ref{fig:DORvsOutage}. Therefore, obtaining DOR requires an accurate approximation of the tail of the CDF of SNR. 

\subsubsection{Contributions}

We focus on a URLLC-IoT system, where a cluster of IoT devices collaborates to perform reliable low-latency uplink transmission towards an Information Collecting Unit (ICU), which can be either a fixed node or, for example, a drone. The primary emphasis is on scenarios involving Rician fading, where a LoS link is established between the transmitting devices and the receiver in the ICU. We consider two scenarios: In the first scenario, information from a CKM is used to roughly adjust the phases of the transmitting IoT devices. In the second scenario, IoT devices employ quantized channel side-information provided by the ICU receiver. The main contributions are: 
\begin{itemize}
\item We propose a cooperative transmission model designed to address scenarios with either limited receiver locations or inaccurate channel side-information in the transmitting devices.
\item The calculations of the DOR and outage probability are derived under the assumption of Rician fading, enabling a comprehensive performance evaluation ranging from static channels with a clear Line of Sight (LoS) to Rayleigh fading cases where a dominant channel direction is absent.
\item The study explicitly demonstrates the impact of reduced side-information quality in the transmission cluster on the cooperative IoT transmission performance. Our findings suggest that performance degradation due to erroneous side-information may be tolerable in terms of outage probability but can be critical for DOR performance.
\item We develop analytical expressions that establish a clear relationship between the DOR, the required number of cooperative IoT devices for successful transmission, and the influence of side-information errors.
\end{itemize}
The rest of the paper is organized as follows: The IoT system model and its protocol are described in Section II, including two different scenarios which are CKM-based coordination and feedback-based coordination, respectively. Then, the DOR expression and outage probability have been derived in Section III. The numerical results are presented and analyzed in Section IV. Finally, Section V concludes the paper. 

\section{IoT System model and protocol}

\begin{figure}[b!]
\centering
\includegraphics[width=9cm]{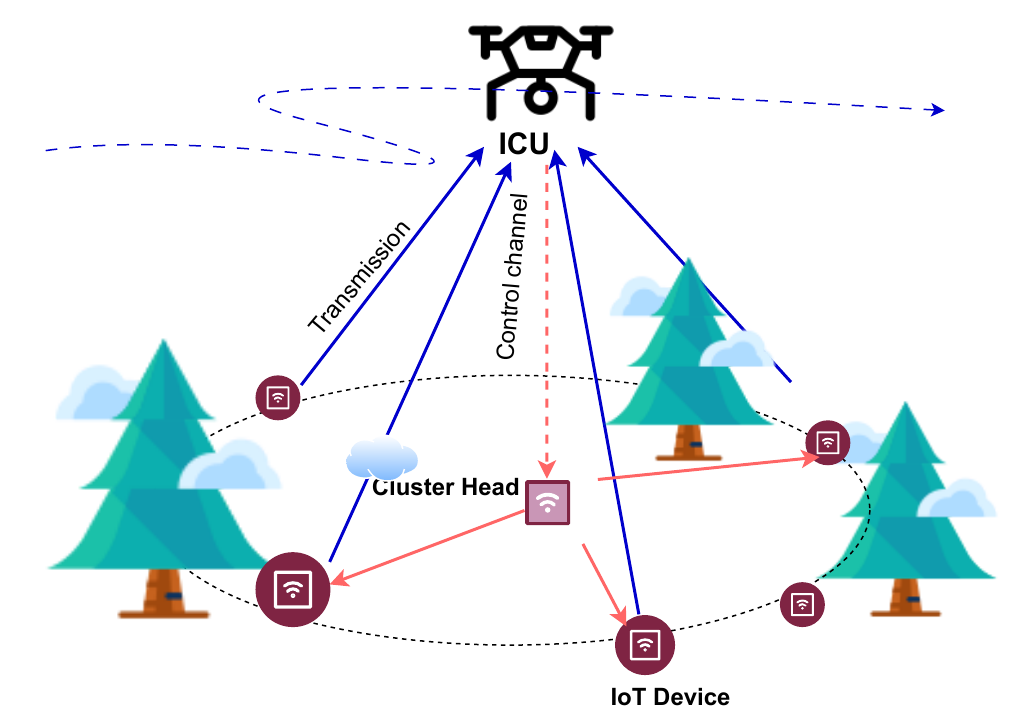}
\caption{Illustration of IoT system: IoT devices in a cluster communicate with the ICU, coordinated by a CH.}
\label{fig:system_illustration}
\end{figure}

We examine a system where IoT devices are organized into a cluster, communicating with a transceiver unit that gathers information from the cluster, as depicted in Fig.\ref{fig:system_illustration}. This transceiver, referred to as the Information Collecting Unit (ICU), can be either a drone or a stationary radio access point with the capability to utilize, store, or forward the collected information. The primary focus of the study is on the synchronized data communication from the IoT device cluster to the ICU. The coordination of IoT devices within the cluster is managed by a cluster head (CH), highlighted in red in Fig. \ref{fig:system_illustration}. The CH engages in control information exchange with the ICU, while the IoT devices handle data transmission. The transmitted data packet remains consistent across all IoT devices.   


\begin{figure}[t!]
\centering
\includegraphics[width=9cm]{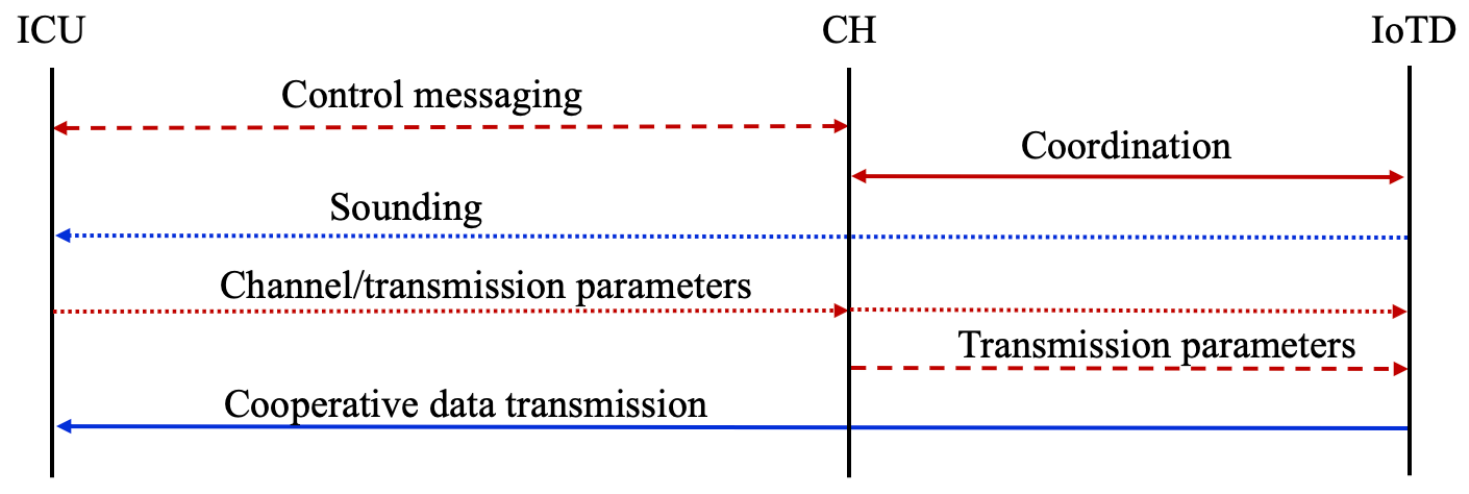}
\caption{Illustration of the communication protocol depicting the coordination phase, where the CH prepares selected IoT devices for transmission and engages in control messaging with the ICU.}
\label{fig:protocol_description}
\end{figure}

The communication procedure is depicted in Fig. \ref{fig:protocol_description}. In this illustration, we consider a scenario where the CH predominantly handles or entirely manages the control messaging with the ICU. Such messaging encompasses various aspects, including admission control, transmission timing information, and other relevant details. These messages serve the purpose of preparing for a communication session or facilitating maintenance for a more prolonged communication between the IoT device cluster and the ICU. During the coordination phase, the CH locally readies specific IoT devices for transmission, identifying those that will participate in the coordinated transmission. The selection criteria for IoT devices may be based on factors such as available power resources in different devices. 

After completing the initial steps, we consider two scenarios for the coordinated transmission:  

\vspace{0.2cm}

\noindent {\bf Scenario 1: CKM-based coordination.}  In this scenario we assume that CKM is available, with mapping $(Z_{\text{ICU}},Z_{\text{IoTD}})\rightarrow\theta$, where $Z_{\text{ICU}}$ and $Z_{\text{IoTD}}$ include the ICU and IoT device coordinates, respectively, and $\theta$ is a phase to be used in the transmission. That is, IoT device apply the weight $e^{j\theta}$ in the transmission to ensure in ICU receiver a coherent sum of signals from different IoT devices. The CKM can be available either in the CH or in IoT devices. In the first option CH carries the burden of storing and maintaining the CKM but, on the other hand, it only needs to inform IoT devices on the applied transmission phases. In the second option IoT device needs to store the CKM of its own but it can choose the phase based on the ICU location. In either of the cases there is no need for channel sounding or CSI feedback from ICU, so at least two stages of the protocol in Fig.\ref{fig:protocol_description} become redundant. Those stages are denoted by dotted arrows.           

\noindent {\bf Scenario 2: Feedback-based coordination.} The CKM-based coordination requires only the location of ICU  due to the availability of CKM either in IoT devices (individual CKM) or in the cluster head (joint CKM). As noted, location based phasing may work only in case of strong LoS. In case of weak LoS coherent signal combining over IoT devices may take place only if there is timely phasing information available in the transmitting devices. Such information can be based on the quantized phase feedback from the ICU. Accordingly, we assume in this scenario that the ICU sends a quantized phase information per IoT device. This phase information is either sent directly to each IoT device or to the CH which shares the information with IoT devices  and organizes locally the cooperative transmission later on. In both options the phase information should contain only very few bits since the feedback channel capacity is limited and amount of bits to fed back is increasing with number of IoT devices.

\section{Signal, channel and side-information models}
This section introduces models for signals, channels, and side-information which will be used in the analysis later on. 
\subsection{Signal and channel models}
Consider a signal model $r=\langle{\bf h},{\bf w}\rangle s+n$, where ${\bf h}=(h_{k})_{k=1}^K\in\mathbb{C}^{K}$ contains complex channel coefficients. Each IoT device admits a fixed transmission power $P$ and ${\bf w}=(\sqrt{P}\delta_ke^{j\phi_k})_{k=1}^K$, where $\delta_k=1$ when $k^{th}$ device is active and $\delta_k=0$ otherwise. Thus, the total transmission power $\sum_{k=1}^KP\delta_k$ applied in the cluster of devices increases with the number of transmitting devices unless we set $P=P_0/|\delta|$. By $\phi_k$ we denote the phase applied in the $k^{th}$ device based on the side-information. Finally, $s$ refers to the transmitted symbol and $n$ represents a complex AWGN term.  

The following assumptions regarding the channels are considered: 
\begin{itemize}
\item  The average propagation loss (including distance dependent loss, shadowing and antenna gains) is the same for all IoT devices belonged by the same cluster such that $\mathop{\mathbb{E}}\left[|h_k|^2\right]=\bar{\gamma}$ where $\mathop{\mathbb{E}}\left[\cdot\right]$ is the expectation operator.
\item Rician fading is assumed so the channel coefficients are of the form $h_k=g_k+\sqrt{\gamma_d} e^{j\varphi_k}$, where $g_k$ represents the scattered and reflected signal part that is a complex zero-mean Gaussian with mean power $\mathop{\mathbb{E}}\left[|g_k|^2\right]=\bar{\gamma}_s$.  Furthermore, $\gamma_d\geq0$ is the power of the static signal part and the amplitudes $|h_k|=\sqrt{\gamma_k}$ are independent and identically distributed (i.i.d.) Rician random variables with the same Rice factor $\nu=\gamma_d/\bar{\gamma}_s$. 
\item  The phase $\varphi_k$ of the static signal part is a sample from a uniform distribution. Yet, we assume that phase $\varphi_k$ is fixed or phase drift is so small that it can be neglected. 
\end{itemize}

Based on these assumptions, the IoT devices are organized in clusters, where the mean path loss is uniform for all devices in relation to the ICU. This assumption is particularly valid under LoS conditions, as the distance between the ICU and the IoT device cluster is typically much greater than the size of the cluster. Consequently, the path loss and ICU antenna gain remain nearly constant for all devices within the cluster. The choice of Rice fading is deliberate, as it offers versatility in modeling the fading channel. In environments characterized by rich scattering around the ICU or device cluster, the Rice factor may be small. Conversely, in the presence of a strong LoS component, the Rice factor is large, resulting in an almost static channel.
\subsection{Side-information models}
The side-information is related to  the phases $\{\phi_k\}_{k=1}^{K}$. We consider two models for the use of side-information for the CKM-based coordination and the feedback-based coordination, respectively: 

\vspace{0.2cm}

\noindent {\bf CKM-based coordination.} In this scenario IoT devices in the cluster have predefined, location dependent, phase values that they apply once the position of the ICU is known. That is, in each device there is a scarce CKM that defines a mapping between the ICU location and a phase weight. Another implementation option is to store a joint CKM in the cluster head. After receiving/estimating the ICU location the cluster head informs each device regarding the applied phase weight. We assume that CKM is computed or otherwise defined based on pure LoS condition. That is, if the static signal part has the phase $\varphi_k$, then the applied phasing is $\phi_k=-\varphi_k+\epsilon_k$, where $\epsilon_k$ represent the error emerging from the estimation and possible quantisation of the CKM information. Now we have 
\begin{equation}\label{eq:component_rice_channel}
\begin{split}
h_kw_k&=\sqrt{P}(g_k+\sqrt{\gamma_d} e^{j\varphi_k})e^{-j\phi_k}\\
&=\sqrt{P}(g_k+\sqrt{\gamma_d}  e^{j\varphi_k})e^{-j(\varphi_k-\epsilon_k)}\\
&=\sqrt{P}(\tilde{g}_k+\sqrt{\gamma_d}  e^{j\epsilon_k}).
\end{split}
\end{equation} 
We note that phasing does not impact the properties of the complex Gaussian part of the signal. That is, $\tilde{g}_k$ admit the same statistical properties as $g_k$. Furthermore, if the Rice factor is small, then it becomes apparent that the usefulness of CKM information will be small as well. Since the phase error may include contribution from many sources of unidealites, we assume $\epsilon_k\sim{\cal N}(0,\sigma_\epsilon^2)$. That is, error follows the zero-mean Gaussian distribution with variance $\sigma_\epsilon^2$.

\vspace{0.2cm}

\noindent {\bf Feedback-based coordination.} Here we assume that $N$-bit phase information is available in IoT devices. If this phase information is send directly from ICU to each device, then dedicated feedback channels are needed. Another option is that ICU send all phase information ($|\delta|\cdot N$ bits) to the CH that then forwards the $N$ bits of information to each IoT device. We apply a uniform phase quantisation: $\phi_k\in\{n\pi/2^{N-1}:\, n=1,2,\dots,2^N\}$. If the phase information is error-free, then the received signal is given by 
\begin{equation}\label{eq:received_signal}
\langle{\bf h},{\bf w}\rangle=\sqrt{P}\sum_{k=1}^{K}\delta_k\sqrt{\gamma_k}e^{j\theta_k},\quad \theta_k\sim {\cal U}\Big(-\frac{\pi}{2^N},\frac{\pi}{2^N}\Big),
\end{equation}
 where ${\cal U}(-a,a)$ refers to the uniform distribution over the interval $(-a,a)$. However, side-information errors may occur and, in that case, the whole word of $N$ bits including the side-information to a certain device, becomes random. The side-information word error probability is denoted by $p_w$.


\section{Analytic Performance Evaluation}

\subsection{Outage probability and DOR}\label{sec:performance_indicators}

The outage probability can be defined as a probability that link rate falls below a certain predefined minimum level. Assume a transmission bandwidth $W$ and a single data stream transmission. Then we may use the classical Shannon formula $R=W\cdot\log_2(1+\gamma)$ for an instantaneous rate $R$ and 
\begin{equation}\label{eq:outage_probability_definition}
\begin{split}
P_\text{out}&=P(R<R_{\min})\\
&=P(\gamma<2^{R/R_{\min}}-1)=F_\gamma(2^{R/R_{\min}}-1),
\end{split}
\end{equation}   
where $F_\gamma(\cdot)$ is the CDF of SNR. Analyzing outage probability is crucial for evaluating the coverage range of a wireless service with a specified rate requirement $R_{\min}$. The computation of outage relies on the distribution of SNR. While closed-form expressions for SNR distribution exist in classical scenarios with single transmitting devices/antennas, incorporating practical aspects such as errors, quantized side-information, and multiple devices poses challenges, often requiring approximations due to the complexity of the problem.

Conventionally, the outage probability in \eqref{eq:outage_probability_definition} is considered in the 5\% - 95\% range. For classical coverage studies 5\% outage probability is typical while in case of delay tolerant IoT services even 95\% packet loss maybe tolerable if numerous re-transmissions can be used, as in narrow band (NB) IoT system \cite{Wang2017,AbRa2022}. If the communication latency is limited, then it is favorable to avoid re-transmissions by using coordinated transmission and apply DOR to measure the link latency. While the CDF of SNR can be used in the computation of DOR as in case of classical coverage studies, there is a an important practical difference. Namely, the left tail of the SNR distribution is in a key role and if approximation is used for the CDF, then the approximation error easily becomes the limiting factor in the DOR assessment.    

The DOR is defined as the probability that the
required information delivery time for a specific transmission session exceeds the predefined threshold \cite{YaCh2020}. Assume an ideal continuous-rate adaptive transmission, where transmitter always transmits at the maximum transmission rate that the channel can reliably support. In such optimal rate adaptation transmission the instantaneous rate is given by $R=W\cdot\log_2(1+\gamma)$. Furthermore, if $D$ bits are transmitted, then the data delivery time $T_d$ is given by $T_d=D/R$. Following the discussion in \cite{YaCh2020} we define the DOR as the probability 
\begin{equation}\label{eq:DOR_definition}
P_{\text{dor}}=P(T_d>T_{th})=P(R<D/T_{th}),
\end{equation}
where $T_{th}$ is time delay threshold for the transmission. Thus, we obtain  
 \begin{equation}\label{eq:DOR_SNR}
P_{\text{dor}}=P\big(\gamma<2^{D/WT_{th}}-1\big)=F_{\gamma}\big(2^{D/WT_{th}}-1\big).
\end{equation}
As evident from \eqref{eq:DOR_SNR}, $P_{\text{dor}}$ approaches one when $T_{th}$ is very small. However, as $T_{th}$ increases, the fraction $D/WT_{th}$ becomes small, and at a certain point, $P_{\text{dor}}$ decays rapidly. This behavior is indicative, especially in a logarithmic scale, of the threshold $T_{th}$ at which reliable transmission, such as a five-nine transmission, becomes possible. 
At the end, to compute both the outage probability and delay outage rate, the distribution of the SNR seen by the receiver is required. For that purpose we deduce in the following accurate analytic approximations for the SNR distribution in both Scenario 1 and Scenario 2 introduced in Section II.   

\subsection{SNR distribution for CKM scenario}

In the CKM there is a mapping defined between the location of the ICU receiver and the favorable phase in each device. Since the mapping is defined in case of pure LoS, it will impact only the static channel part while the phase of the reflected/scattered part of the signal is random. Using \eqref{eq:component_rice_channel} we write    
\begin{equation}
\begin{split}
\langle{\bf h},{\bf w}\rangle_1&=\sqrt{P}\Big(\sum_{k=1}^{K}\delta_k\sqrt{\gamma_{s,k}}e^{j\psi_k}+\sqrt{\gamma_d}\sum_{k=1}^{K}\delta_ke^{j\epsilon_k}\Big)\\
&=\sqrt{P}\left(g_s+\sqrt{\gamma_{d,\Sigma}}\cdot e^{j\epsilon_{\Sigma}}\right),
\end{split}
\end{equation}
where subscript in $\langle{\bf h},{\bf w}\rangle_1$, '1' ,refers to the Scenario 1. We note that here $g_s$ contains the reflected/scattered signal part and $\gamma_{d,\Sigma}$ is the static part of the sum signal after the phasing defined by CKM is used, and notation $\epsilon_{\Sigma}$ refers to the phase of the static part. Since $g_s$ is composed by complex zero-mean Gaussian variables, it is also itself a complex zero-mean Gaussian with power $\mathop{\mathbb{E}}\left[|g_s|^2\right]=\bar{\gamma}_s|\delta|$. Thus, $|\langle{\bf h},{\bf w}\rangle_1|$ is a Rice variable with Rice factor depending on the phasing errors $\epsilon=(\epsilon_k)_k$. 

In what follows, we approximate the distribution of $|\langle{\bf h},{\bf w}\rangle_1|$ by the Rice distribution with a Rice factor $\nu_{\Sigma}=\mathop{\mathbb{E}}\left[\gamma_{d,\Sigma}\right]/E\mathop{\mathbb{E}}\left[|g_s|^2\right]$. We have   
\begin{equation}
\begin{split}
\bar{\gamma}_{d,\Sigma}&=\mathop{\mathbb{E}}\left[\gamma_{d,\Sigma}\right]=\gamma_d\mathop{\mathbb{E}}\left[\Big|\sum_{k=1}^{K}\delta_ke^{j\epsilon_k}\Big|^2\right]\\
&=\gamma_d\Big(\mathop{\mathbb{E}}\left[\langle\delta,\cos\epsilon\rangle^2\right]+\mathop{\mathbb{E}}\left[\langle\delta,\sin\epsilon\rangle^2\right]\Big)
\end{split}
\end{equation} 

In order to obtain expectations of $\langle\delta,\cos\epsilon\rangle^2$ and $\langle\delta,\sin\epsilon\rangle^2$ we need to compute $\mathop{\mathbb{E}}\left[\cos^n\epsilon_k\right]$ and $\mathop{\mathbb{E}}\left[\sin^n\epsilon_k\right]$for $n=1,2$. Since $\epsilon_k\sim{\cal N}(0,\sigma_\epsilon^2)$ we have 
\begin{equation}\label{eq:phasing_gaussian_error}
\begin{split}
\mathop{\mathbb{E}}\left[\cos \epsilon_k\right]&=\frac{1}{\sqrt{2\pi}\sigma_\epsilon}\int_{-\infty}^{\infty} e^{-\frac{\epsilon_k^2}{2\sigma_\epsilon^2}}\cos\epsilon_k\, d\epsilon_k=e^{-\frac{\sigma_\epsilon^2}{2}},
\end{split}
\end{equation}   
where the second equality follows by using formula (3.896.2) of \cite{GrRy2007}. Function $\sin\epsilon_k$ is odd and phase error is symmetric around zero, implying $\mathop{\mathbb{E}}\left[\sin \epsilon_k\right]=0$. It remains to compute $\mathop{\mathbb{E}}\left[\cos^2 \epsilon_k\right]$ and $\mathop{\mathbb{E}}\left[\sin^2 \epsilon_k\right]$. By using the identities [(3.898.1-2), \cite{GrRy2007}], we get
\begin{equation}\label{eq:gaussian_error2}
\mathop{\mathbb{E}}\left[\cos^2\epsilon_k\right]=\tfrac12(1+e^{-\frac{\sigma_\epsilon^2}{2}}),\ 
\mathop{\mathbb{E}}\left[\sin^2\epsilon_k \right]=\tfrac12(1-e^{-\frac{\sigma_\epsilon^2}{2}}).
\end{equation}  
By expanding $\langle\delta,\cos\epsilon\rangle^2$ we obtain 
\begin{equation}
\begin{split}
\mathop{\mathbb{E}}\left[\langle\delta,\cos\epsilon\rangle^2\right]&=\sum_{k=1}^K\delta_k\mathop{\mathbb{E}}\left[\cos^2\epsilon_k\right]+\dots\\
\dots+\sum_{k=1}^K&\sum_{l\neq k}\delta_k\delta_l \mathop{\mathbb{E}}\left[\cos\epsilon_k\right]\mathop{\mathbb{E}}\left[\cos\epsilon_l\right]\\
=\tfrac12|\delta|\big(1&+e^{-\frac{\sigma_\epsilon^2}{2}}\big)+|\delta|(|\delta|-1)e^{-\sigma_\epsilon^2}.
\end{split}
\end{equation}
Similarly, we get 
\begin{equation}
\begin{split}
\mathop{\mathbb{E}}\left[\langle\delta,\sin\epsilon\rangle^2\right]&=\sum_{k=1}^K\delta_k\mathop{\mathbb{E}}\left[\sin^2\epsilon_k\right]
=\tfrac12|\delta|\big(1-e^{-\frac{\sigma_\epsilon^2}{2}}\big).
\end{split}
\end{equation}
The Rice factor of the approximation attains the form
\begin{equation}\label{eq:rice_factor_sum}
\begin{split}
&\nu_\Sigma=\frac{\bar{\gamma}_{d,\Sigma}}{\bar{\gamma}_s|\delta|}=\nu\big(1+(|\delta|-1)e^{-\sigma_\epsilon^2}\big).\\
\end{split}
\end{equation}
We note that Rice factor increases with the number of devices cooperating but phase errors may crucially impact the resulting  performance gain. The approximate CDF of SNR obtained from Rice distribution is given in terms of the Marqum Q-function:
\begin{equation}\label{eq:CDF_scenario1}
F_{\hat{\gamma}}(\gamma)=1-Q_1\Big(\tfrac{\sqrt{\bar{\gamma}_{d,\Sigma}}}{\sigma_{\Sigma}},\tfrac{\sqrt{\gamma}}{\sigma_{\Sigma}}\Big),\  
\sigma_{\Sigma}=\sqrt{\tfrac12|\delta|\bar{\gamma}_s},
\end{equation}
where $\hat{\gamma}$ refers to the approximation of $|\langle{\bf h},{\bf w}\rangle_1|^2$. We further note that the terms $\bar{\gamma}_{d,\Sigma}$ and $\sigma_{\Sigma}$ in \eqref{eq:CDF_scenario1} are functions of $\nu$, $|\delta|$, $\sigma_\epsilon$ and $\bar{\gamma}$.   
Using \eqref{eq:CDF_scenario1} we obtain for the DOR
\begin{equation}\label{eq:DOR_scenario1}
P_{\text{dor}}=1-Q_1\Big(\tfrac{\sqrt{\bar{\gamma}_{d,\Sigma}}}{\sigma_{\Sigma}},\tfrac1{\sigma_{\Sigma}}\small{\sqrt{2^{D/WT_{th}}-1}} \Big).
\end{equation}
That is, for a given data volume $D$ and bandwidth $W$, we can compute the DOR as a function of delay threshold $T_{th}$. 

Now we can also consider the following question: \textit{If, for a certain service, the DOR and delay threshold are given, how many transmitting devices are needed to ensure that service requirements are fulfilled?} 

\noindent It was shown in the Appendix that
\begin{equation}\label{eq:number_of_IoTD_scenario1}
|\delta|> -\tfrac{e^{\sigma_{\epsilon}^2}}{4\nu}\log(2P_{\text{dor}})\left(1+\sqrt{1-
4e^{-\frac{\sigma_{\epsilon}^2}{2}}\tfrac{\sqrt{\nu(1+\nu)\gamma/\bar{\gamma}}}{\log(2P_{\text{dor}})}} \right)^2,
\end{equation}
where the required SNR is $\gamma=2^{D/WT_{th}}-1$ and \eqref{eq:number_of_IoTD_scenario1} is valid only if 
\begin{equation}\label{eq:condition_for_number_eq}
T_{th}>D\big(W\log_2(1+\tfrac{\bar{\gamma}\nu}{1+\nu}|\delta|^2e^{-\sigma_{\epsilon}^2})\big)^{-1}.
\end{equation} 
The bound \eqref{eq:number_of_IoTD_scenario1} provides means to assess the number of required devices in cooperative transmission but it assumes that the mean power $\bar{\gamma}$ and Rice factor $\nu$ of the channel are known and standard deviation $\sigma_\epsilon$ of the errors can be assessed. The first two are related to locations of transmitter and receiver, and can be obtained from CKM. The standard deviation of error can be also predefined when CKM is formed.    

If the total transmission power in cluster is kept constant, i.e. the transmission power in IoT devices is scaled by number of devices, then we have shown in Appendix that the requirement for the number of devices attains the form
\begin{equation}\label{eq:number_of_IoTD_scenario1_scaled}
|\delta|> e^{\sigma_{\epsilon}^2}\Big(\sqrt{\tfrac1{\nu}\log(2P_{\text{dor}})^{-1}}+\sqrt{\tfrac{1+\nu}{\nu}\tfrac{\gamma}{\bar{\gamma}}}\Big)^2.
\end{equation}

\subsection{SNR distribution when using channel side-information}

In this case, the sum channel is composed of complex components that are adjusted based on the side-information from the receiver.  Therefore, we adopt approximation that is based on the central limit theorem. Let us start by writing the SNR of the received signal in the form  
\begin{equation}\label{eq:received_SNR}
\gamma=\big|\langle{\bf h},{\bf w}\rangle_2\big|^2=(\text{Re}\langle{\bf h},{\bf w}\rangle_2)^2+(\text{Im}\langle{\bf h},{\bf w}\rangle_2)^2,
\end{equation}
where subscript '2' refers to the Scenario 2 and both $\text{Re}\langle{\bf h},{\bf w}\rangle_2$ and $\text{Im}\langle{\bf h},{\bf w}\rangle_2$ are composed of signals from $|\delta|$ devices. If the number of devices is large, then we can use the central limit theorem to approximate both real and imaginary parts of $\langle{\bf h},{\bf w}\rangle_2$ by Gaussian variables \cite{HaHa2013}. More precisely, we approximate $\text{Re}\langle{\bf h},{\bf w}\rangle_2\approx X$,  $\text{Im}\langle{\bf h},{\bf w}\rangle_2\approx Y$ and $\gamma\approx\hat{\gamma}=X^2+Y^2$, where $X$ and $Y$ are Gaussian with the following means and variances:
\begin{equation}\label{eq:parameters}
\begin{cases}
\mu_r=\mathop{\mathbb{E}}\left[\text{Re}\langle{\bf h},{\bf w}\rangle_2\right],\ \sigma_r^2=\mathop{\mathbb{E}}\left[(\text{Re}\langle{\bf h},{\bf w}\rangle_2)^2\right]-\mu_r^2,\\
\mu_i=\mathop{\mathbb{E}}\left[\text{Im}\langle{\bf h},{\bf w}\rangle_2\right],\ \sigma_i^2=\mathop{\mathbb{E}}\left[(\text{Im}\langle{\bf h},{\bf w}\rangle_2)^2\right]-\mu_i^2. \\
\end{cases}
\end{equation}
Here we note that the distribution of variable $\theta_k$ in \eqref{eq:received_signal} is symmetric around zero and $\sin\theta_k$ is an odd function. Therefore $\mu_i=0$ and $Y^2$ follows the central $\chi^2$-distribution with one degree of freedom. On the other hand, $\mu_r\neq0$ and $X^2$ follow the non-central  $\chi^2$-distribution with one degree of freedom. The related non-centrality parameter of this distribution is $\lambda=\mu_r^2$.

From the classical theory of statistics, it is known that the sum $\hat{\gamma}=X^2+Y^2$ follows the $\chi^2$-distribution with two degrees of freedom. Moreover, the related PDF and CDF can be given in the following form:
\begin{equation}\label{pdfwk}
\begin{split}
f_{\hat{\gamma}}(\gamma)&=\sum_{n=0}^\infty a_nf_n(\gamma),\ F_{\hat{\gamma}}(\gamma)=\sum_{n=0}^\infty a_nF_n(\gamma),\\
\end{split}
\end{equation}   
where the weights $a_n$ are defined as
\begin{equation}\label{weight_factor}
a_n=\tfrac{\Gamma(n+\frac12)}{\Gamma(n+1)\Gamma(\frac12)}\left(\tfrac{\sigma_r}{\sigma_i}\right)\left(\tfrac{\sigma_i^2-\sigma_r^2}{\sigma_i^2}\right)^n
\end{equation}
and the component functions $f_n$ and $F_n$ are of the form:
\begin{equation}\label{pdf_cdf}
\begin{split}
f_n(\gamma)&=\tfrac{1}{2\sigma_r^2}\left(\tfrac{\gamma}{\lambda}\right)^{\frac{n}{2}}e^{-\frac{\lambda+\gamma}{2\sigma_r^2}}I_n\left(\tfrac{\sqrt{\lambda}}{\sigma_r^2}\sqrt{\gamma}\right),\\
F_n(\gamma)&=1-Q_{n+1}\left(\tfrac{\sqrt{\lambda}}{\sigma_r},\tfrac{\sqrt{\gamma}}{\sigma_r}\right),\\
\end{split}
\end{equation}
where $I_n(\cdot)$ is the modified Bessel function of order $n$ and $Q_m(\cdot,\cdot)$ is the generalised Marcum Q-function of order $m$.

Before computing the parameters \eqref{eq:parameters}, we consider the impact of an erroneous side-information. The side-information is composed by $|\delta|$ $N$-bit words, each word dedicated to a certain device. If we assume that the word error probability is $p_w$ and errors occur independently, then 
\begin{equation}\label{eq:scenario2_CDF_errors}
\begin{split}
F_{\hat{\gamma}}(\gamma)=\sum_{m=0}^{|\delta|}\binom{|\delta|}{m}(1-p_w)^{|\delta|-m}p_w^m F_{\hat{\gamma}}(\gamma |m),\\
\end{split}
\end{equation}   
where $F_{\hat{\gamma}}(\gamma |m)$ refers to the CDF of SNR in case there are $m$ erroneous phasing words. We may use the formulas \eqref{pdfwk}-\eqref{pdf_cdf} to compute $F_{\hat{\gamma}}(\gamma |m)$, but it is important to notice that parameters defined in \eqref{eq:parameters} will depend on the number $m$ of erroneous phasing words. Therefore, the definitions in  \eqref{eq:parameters} for the error-free case and for the case  where the phasing word error exists should be computed.   

Let us now deduce closed-form expressions for the parameters \eqref{eq:parameters}. We have 
\begin{equation}\label{eq:received_SNR_scenario2}
\langle{\bf h},{\bf w}\rangle_2=\sqrt{P}\cdot\sum_{k=1}^{K}\delta_k\sqrt{\gamma_k}e^{j\theta_k}.
\end{equation}
To find the required expressions, we need expectations of the first and second power of real and imaginary parts of $\langle{\bf h},{\bf w}\rangle_2$. 
First, to obtain $\mu_{r}$ we compute $\mathop{\mathbb{E}}\left[\text{Re}\langle{\bf h},{\bf w}\rangle_2\right]$. For that purpose we recall that the expected value for the amplitude of a Rician variable is given by 
\begin{equation}\label{eq:Rice_Amplitude_expectation}
\begin{split}
\mathop{\mathbb{E}}\left[\sqrt{\gamma_k}\right]&=\sqrt{\tfrac{\pi\bar{\gamma}}{4(1+\nu)}}e^{-\frac{\nu}2}\left((1+\nu)I_0(\tfrac{\nu}2)+\nu I_1(\tfrac{\nu}2)\right).\\
\end{split}
\end{equation}  
Moreover, in the error-free case $\theta_k\sim{\cal U}\left(-\frac{\pi}{2^{N}},\frac{\pi}{2^{N}}\right)$ and we obtain 
\begin{equation}\label{eq:phasing_expectation}
\mathop{\mathbb{E}}\left[\cos\theta_k\right]=\frac{2^{N-1}}{\pi}\int_{-\frac{\pi}{2^{N}}}^{\frac{\pi}{2^{N}}}\cos\theta\,d\theta=\text{sinc}(2^{-N}),
\end{equation}  
where $\text{sinc}(x)=\sin (\pi x)/(\pi x)$. If error occurs, then $\theta_k\sim{\cal U}\left(-\pi,\pi\right)$ and $\mathop{\mathbb{E}}\left[\cos\theta_k\right]=0$. 

Assume that there are $m$ erroneous phasing words. Then, using equations \eqref{eq:received_SNR_scenario2}-\eqref{eq:phasing_expectation} we obtain
\begin{equation}
\begin{split}
\mu_{r}(m)&=\sqrt{P}\cdot(|\delta|-m)\cdot\sqrt{\tfrac{\pi\bar{\gamma}}{4(1+\nu)}}e^{-\frac{\nu}2}\big((1+\nu)I_0(\tfrac{\nu}2)+\\&\dots+\nu I_1(\tfrac{\nu}2)\big)
\text{sinc}(2^{-N})
\end{split}
\end{equation}
where $1\leq m\leq|\delta|$. Since the $\sin\theta_k$ is odd function, $\mathop{\mathbb{E}}\left[\sin\theta_k\right]=0$ whether $\theta_k\sim{\cal U}\left(-\frac{\pi}{2^{N}},\frac{\pi}{2^{N}}\right)$  or $\theta_k\sim{\cal U}\left(-\pi,\pi\right)$. Therefore $\mu_{i}(m)=0$ for all $m$.

Next we compute the expectations of $\left(\text{Re}\langle{\bf h},{\bf w}\rangle_2\right)^2$ and $\left(\text{Im}\langle{\bf h},{\bf w}\rangle_2\right)^2$.
For the former we have 
\begin{equation}\label{eq:square_formula}
\left(\text{Re}\langle{\bf h},{\bf w}\rangle_2\right)^2=P\sum_{k=1}^{K}\sum_{k=1}^{K}\delta_k\delta_l\sqrt{\gamma_k}\sqrt{\gamma_l}\cos\theta_k\cos\theta_l
\end{equation}  
and thus, we need the following expectations: $\mathop{\mathbb{E}}\left[\sqrt{\gamma_k}\right]$, $\mathop{\mathbb{E}}\left[\cos\theta_k\right]$, $\mathop{\mathbb{E}}\left[\gamma_k\right]$ and $\mathop{\mathbb{E}}\left[\cos^2\theta_k\right]$. The first two were already obtained, so it remains to compute $\mathop{\mathbb{E}}\left[\cos^2\theta_k\right]$. If phasing word is correct, then $\theta_k\sim{\cal U}\left(-\frac{\pi}{2^{N}},\frac{\pi}{2^{N}}\right)$, and we find that 
\begin{equation}\label{eq:phasing_expectation2}
\begin{split}
\mathop{\mathbb{E}}\left[\cos^2\theta_k\right]&=\tfrac12(1+\text{sinc}(2^{-(N-1)})).\\
\end{split}
\end{equation}  
On the other hand, if phasing word is random, then $\theta_k\sim{\cal U}\left(-\pi,\pi\right)$, and we obtain $\mathop{\mathbb{E}}\left[\cos^2\theta_k\right]=1/2$. By taking the expectation of \eqref{eq:square_formula} we obtain after some elementary manipulations:
\begin{equation}\label{eq:real_square_formula}
\small
\begin{split}
\mathop{\mathbb{E}}\left[\left(\text{Re}\langle{\bf h},{\bf w}\rangle_2\right)^2\right]&=P\big(\tfrac12 m \bar{\gamma}+\tfrac12(|\delta|-m)\bar{\gamma}(1+\text{sinc}(\tfrac{1}{2^{N-1}}))\\
+(|\delta|-m)&(|\delta|-m-1)\mathop{\mathbb{E}}\left[\sqrt{\gamma_k}\right]^2\text{sinc}(\tfrac1{2^{N}})^2\big).
\end{split}
\normalsize
\end{equation}  
As in \eqref{eq:phasing_expectation2}, we have $\mathop{\mathbb{E}}\left[\sin^2\theta_k\right]=\tfrac12(1-\text{sinc}(2^{-(N-1)}))$ and 
\begin{equation}\label{eq:imag_square_formula}
\small
\begin{split}
\mathop{\mathbb{E}}\left[\left(\text{Im}\langle{\bf h},{\bf w}\rangle_2\right)^2\right]&=P\big(\tfrac12 m \bar{\gamma}+\tfrac12(|\delta|-m)\bar{\gamma}(1-\text{sinc}(\tfrac{1}{2^{N-1}})).\\
\end{split}
\normalsize
\end{equation}  
The parameters $\sigma_r(m)$ and $\sigma_i(m)$, $1\leq m\leq |\delta|$ are obtained using \eqref{eq:real_square_formula}, \eqref{eq:imag_square_formula} and the previously computed $\mu_r$ and $\mu_i$.    

 \section{Performance results}

In this section, we present performance results for both Scenario 1 and 2, utilizing two performance indicators. Initially,  the outage probability is investigated for both scenarios. Subsequently, we delve into the results for the delay outage rate, as discussed in Section \ref{sec:performance_indicators}. Although both the outage probability and the delay outage rate employ the same CDF approximation, it's noteworthy that the latter typically focuses on the tail of the distribution, necessitating a highly accurate approximation for the SNR distribution. In contrast, the classical outage probability is plotted on a linear scale, where values and errors of the approximation are discernible around the mean. To assess the precision of our analytical results, we include simulated point values in each figure.

\subsection{Simulation parameters}

To begin with, the mean received SNR (denoted by $\bar{\gamma}$) of an individual device transmission is expected to be low. While the value we apply in performance evaluations is $-15$dB, it could be lower especially for the CKM based phasing of Scenario 1. Yet, if $\bar{\gamma}$ is very small, then even scarce estimation of channel phase in the receiver becomes difficult, making the side-information based approach of Scenario 2 challenging. 

We have scaled the transmission power in the cluster by the number of transmitting devices. This makes easier assessing the gain from cooperation when power gain from additional IoT devices is ignored. Yet, we recall that, for example, with 20 IoT devices the power benefit would be 13dB when compare to a single device transmission. Furthermore, since the focus of this work is on the system that is composed of large number of cooperating IoT devices, we have set the number of cooperating devices to 20 in most performance illustrations. 



In Scenario 1, we assume that phasing error is composed by factors that jointly make the error $\epsilon$ follow the zero-mean Gaussian distribution with standard deviation $\sigma_\epsilon$. Furthermore, we expect that only only in very rare occasions phasing errors lead to destructive summation of static signal components. This expectation is consistent with the assumption $\sigma_\epsilon<30^{\circ}$ because then $|\epsilon|<3\sigma_\epsilon<90^{\circ}$ in almost all cases.   
The number of information bits in each per-device phasing word is either one or two. Using three bits would improve results slightly but focus is more on the use of scarce phase information, since accurate phase estimation of weak device signal is challenging. Finally, phasing word error probability can be high if channel coding is not used.  

Finally, we note that in the DOR performance evaluations in Figs. \ref{fig:delay_outage_gauss_error} and \ref{fig:delay_outage_fb_error} we have set $W=200$kHz and $D=100$bits. These parameters are compatible with e.g. NB-IoT system \cite{Wang2017}. Even though NB-IoT is typically used for delay tolerant services, it is of great interest to understand how to enable low-latency services in similar IoT systems.

 \subsection{CKM based phasing (Scenario 1)}

\begin{figure}
\centering
\includegraphics[width=9cm]{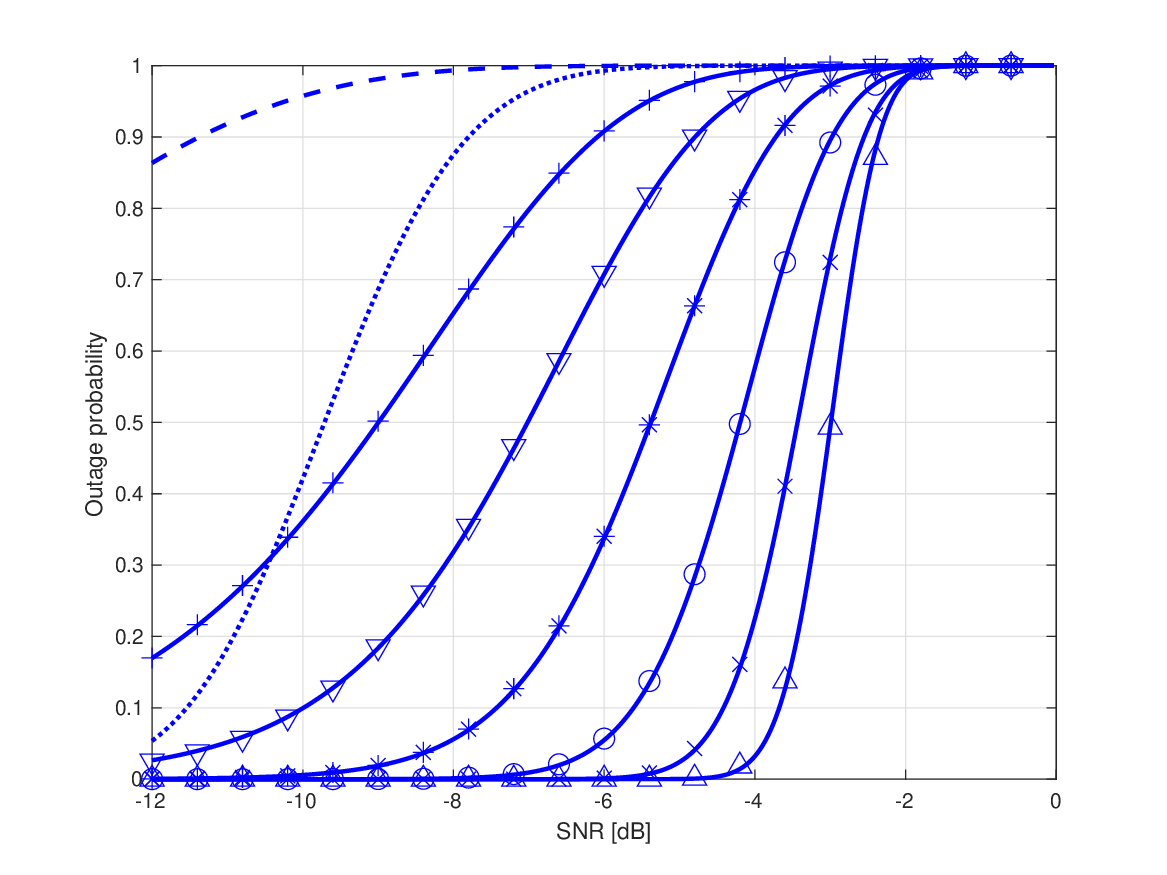}
\caption{Outage probability for CKM-based phasing when $|\delta|=20$ and Gaussian phase error with standard deviation $\sigma_\epsilon=20^{\circ}$ is assumed. Curves: $\nu=-6$dB (+), $\nu=-3$dB ($\nabla$), $\nu=0$dB (*), $\nu=3$dB (o), $\nu=6$dB (x), $\nu=9$dB ($\Delta$). Ticks refer to simulated values and solid curves are plotted using analytic formulae. The dotted curve refers to the outage probability when selection diversity over 20 devices is applied and dashed curve refers to the Rayleigh fading case. }
\label{fig:outage_probability_gauss_error}
\end{figure}

In Fig. \ref{fig:outage_probability_gauss_error} we have the outage probability curves as a function of SNR when $20$ clustered IoT devices cooperate in transmission. The CKM-based phase information is incomplete and phase error $\epsilon$ follows the Gaussian distribution with standard deviation $\sigma_\epsilon=20$ degrees. As can be expected, the strength of the Rice factor strongly impacts on the performance. When there is no static channel component the channel becomes Rayleigh and outage performance is very poor, as can be seen from part of the dashed curve that is visible. However, it is somewhat surprising that cooperative transmission is already providing some gain when $\nu=-6dB$. Furthermore, the performance difference on 10\% outage level between $\nu=0$dB and $\nu=9$dB is 4dB while the difference between $\nu=-6$dB and $\nu=9$dB is around 10dB. We also note that the performance difference is especially large in low outage levels. The outage probability performance of the selection diversity with 20 diversity branches is presented by the dotted curve in Fig. \ref{fig:outage_probability_gauss_error}. The performance of selection diversity is clearly inferior when compared to CKM based approach. More importantly, the power gain that is obtained from multiple devices transmitting simultaneously cannot be achieved in selection diversity, since transmission power of an individual device cannot be scaled up.          

\begin{figure}
\centering
\includegraphics[width=9cm]{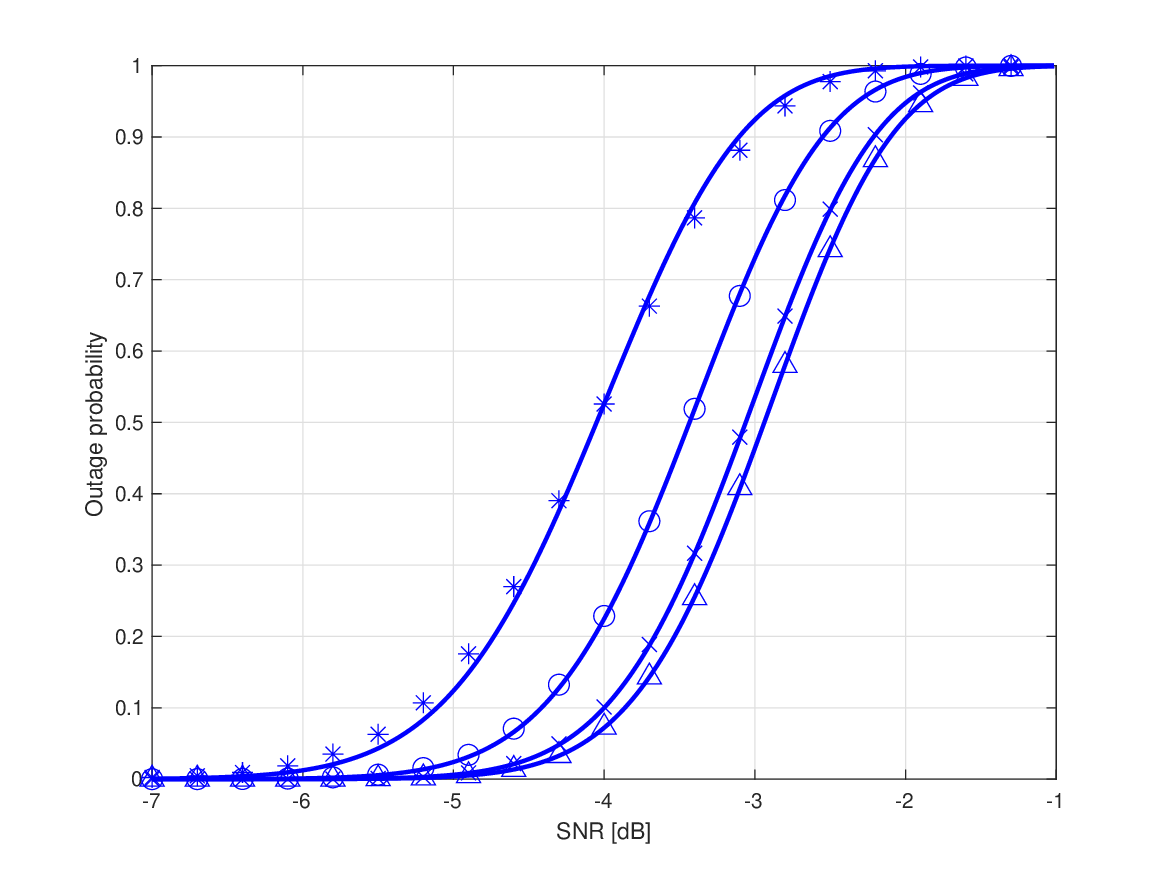}
\caption{Outage probability for CKM-based phasing when $|\delta|=20$, $\nu=6$dB and $\sigma_\epsilon=1^{\circ}$ ($\Delta$), $\sigma_\epsilon=10^{\circ}$ (x), $\sigma_\epsilon=20^{\circ}$ (o) and $\sigma_\epsilon=30^{\circ}$ (*). Ticks refer to simulated values and solid curves are plotted using analytic formulae. }
\label{fig:outage_probability_for_different_errors}
\end{figure}

The impact of Gaussian phasing error on the outage performance of cooperative transmission is illustrated in Fig. \ref{fig:outage_probability_for_different_errors} assuming 6dB Rice factor. Increasing the standard deviation of the error from one degree to 10 degrees has only a minor impact on the performance. However, if it increases to 30 degrees, then the performance loss becomes well visible. We also note that our approximation works best when standard deviation of the error is less than 30 degrees. For larger standard deviation the number of occasions where static components of some IoT devices admit destructive summation increases, degrading both the system performance and the accuracy of the approximation. As can be seen in Fig. \ref{fig:outage_probability_for_different_errors}, there is already some discrepancy between simulated and analytical results when $\sigma_\epsilon=30$ degrees.     

\begin{figure}
\centering
\includegraphics[width=9cm]{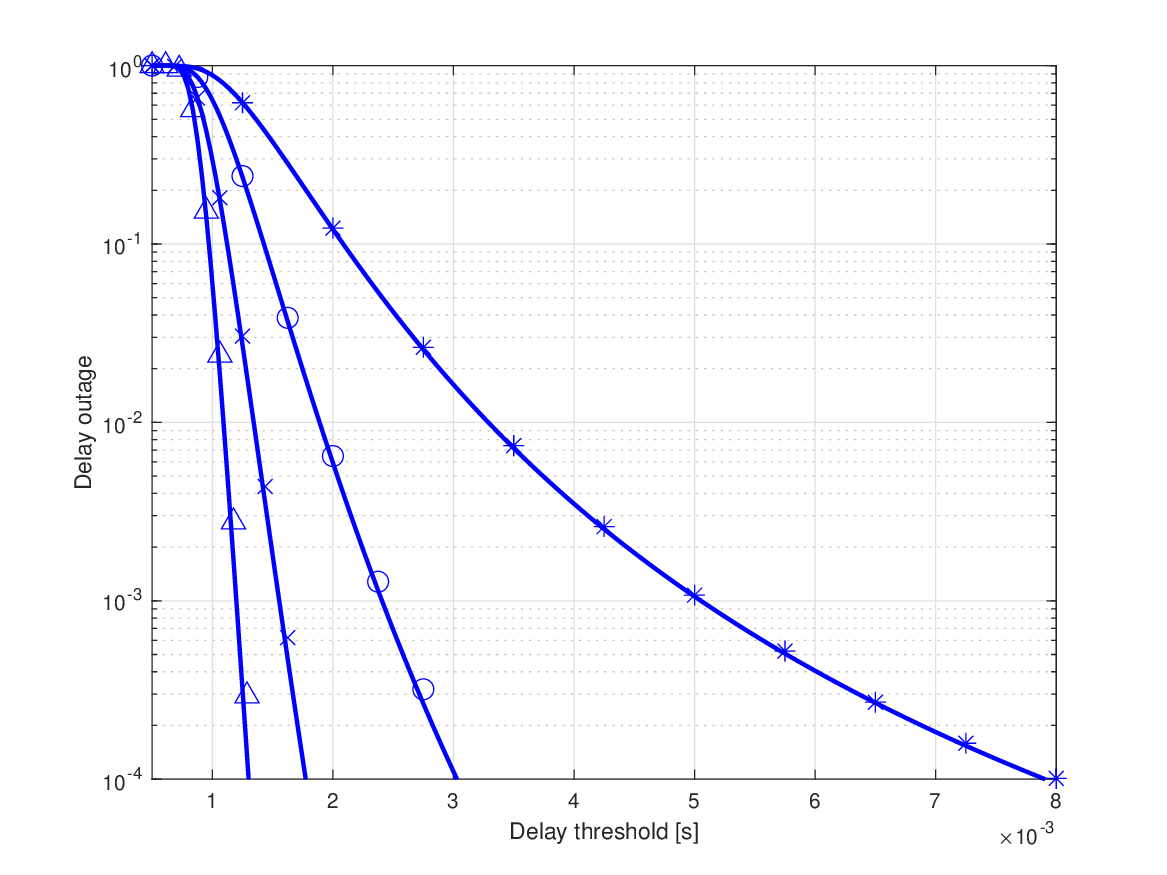}
\caption{Delay outage rate for CKM-based phasing when $|\delta|=20$ and Gaussian phase error with standard deviation $\sigma_\epsilon=20^{\circ}$ is assumed. Curves: $\nu=0$dB (*), $\nu=3$dB (o), $\nu=6$dB (x), $\nu=9$dB ($\Delta$). Ticks refer to simulated values and solid curves are plotted using analytic formulae. }
\label{fig:delay_outage_gauss_error}
\end{figure}

The DOR performance results has been presented in Fig. \ref{fig:delay_outage_gauss_error}. While Fig. \ref{fig:outage_probability_gauss_error} indicated that a Rice factor decrease (e.g., from $\nu=6$dB to $\nu=0$dB) leads to small SNR losses in probability levels between 10\% and 90\%, Fig. \ref{fig:delay_outage_gauss_error} shows that such decrease might be crucial for the delay threshold. If the DOR requirement is e.g. $10^{-4}$ or lower, then even small change in the Rice factor may seriously degrade the quality of service. Thus, low-latency services are very sensitive for any changes in the radio environment. Finally, we note that our approximation works well in the probability range of Fig. \ref{fig:delay_outage_gauss_error}. However, if the probability range is extended to very low probabilities, the approximation errors start to become more visible.     

\begin{figure}
\centering
\includegraphics[width=9cm]{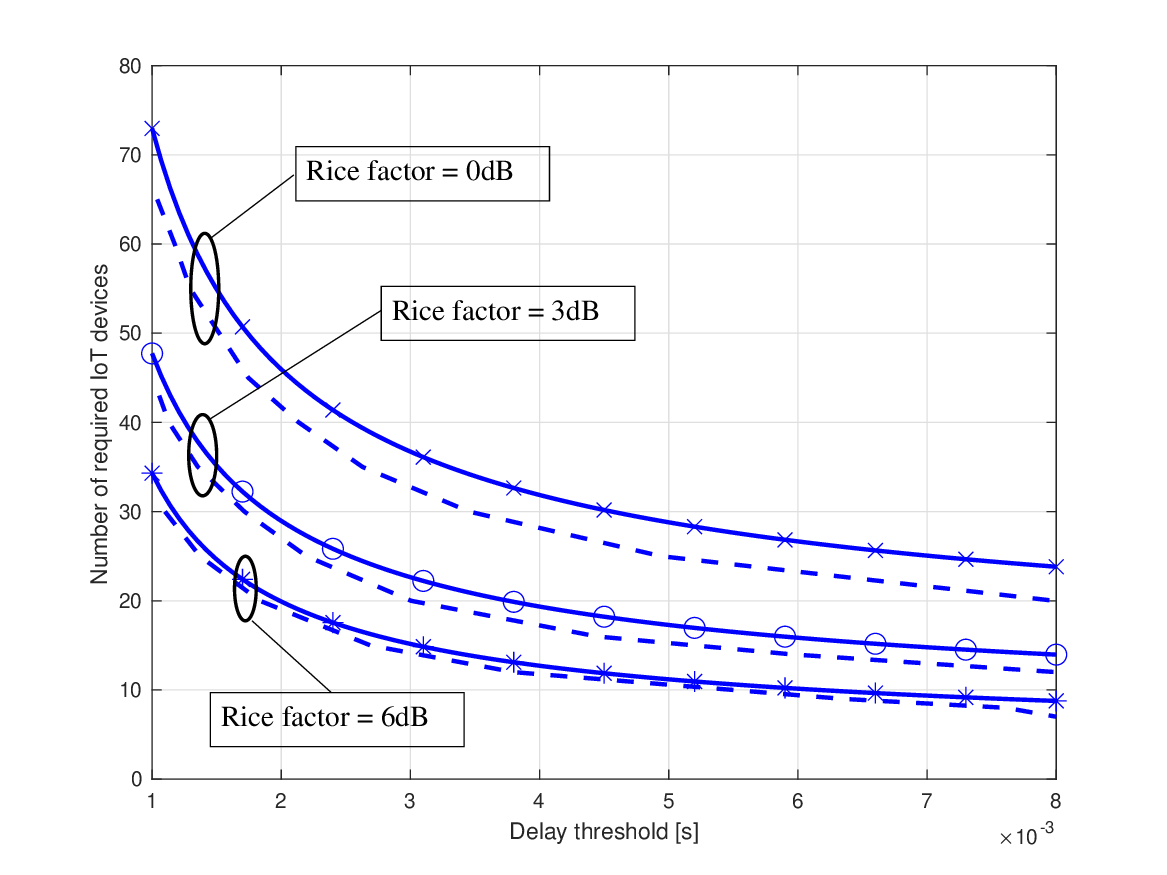}
\caption{Numbers of required IoT devices in CKM-based phasing when $P_{dor}=10^{-4}$ and Gaussian phase error is assumed with standard deviation $\sigma_\epsilon=20^{\circ}$. Rice factors: $\nu=0$dB (x), $\nu=3$dB (o) and $\nu=6$dB (*). Solid curves represent the bounds \eqref{eq:number_of_IoTD_scenario1_scaled} and dashed curves are obtained through simulations. }
\label{fig:number_of_IoTDs_gauss_error}
\end{figure}

The number of required IoT devices in cooperative transmission is plotted in Fig. \ref{fig:number_of_IoTDs_gauss_error} using the analytic bound \eqref{eq:number_of_IoTD_scenario1_scaled} and assuming that $P_{dor}=10^{-4}$. The dashed curves have been obtained through simulations and they indicate that analytic bounds are relatively tight. As can be expected, the strength of the Rice factor has a notable impact on the number of required IoT devices in the cooperative transmission. Namely, if Rice factor is 6dB, then the number of required IoT devices is less than half when compared to the case with 0dB Rice factor. We have shown results only for the case $\sigma_{\epsilon}=20^{\circ}$, but we recall that the impact of phase error is relatively small when $0< \sigma_{\epsilon}\leq 20^{\circ}$.        

 \subsection{Side-information based phasing (Scenario 2)}

\begin{figure}
\centering
\includegraphics[width=9cm]{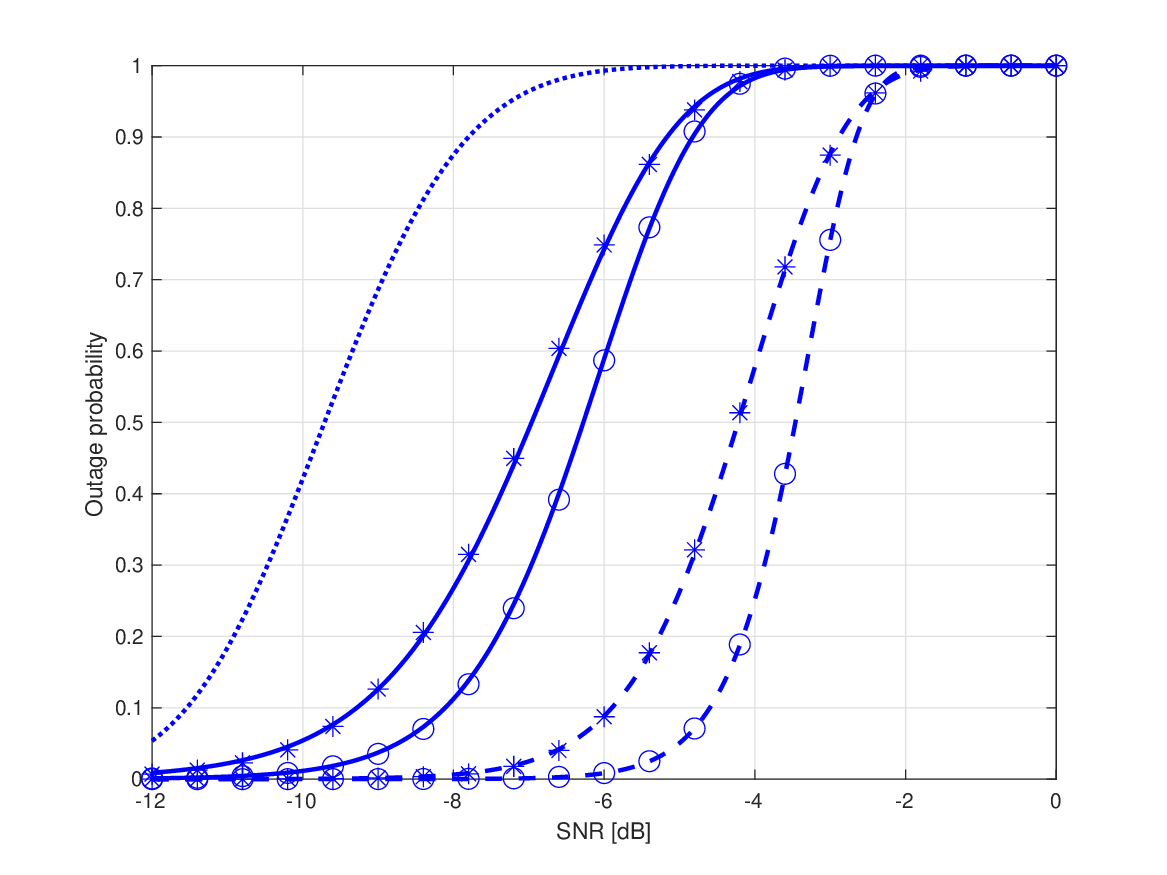}
\caption{Outage probability for side-information based phasing when $|\delta|=20$ and $p_w=0.05$. Solid curves: $N=1$, dashed curves: $N=2$. Rice factors are $\nu=-3$dB (*), $\nu=9$dB (o). Ticks refer to simulated values and solid curves are plotted using analytic formulae. The dotted curve refers to the outage probability when selection diversity over 20 devices is applied. }
\label{fig:outage_probability_fb_error}
\end{figure}

The outage probability results for channel side-information based approach are presented in Fig. \ref{fig:outage_probability_fb_error} assuming cooperative transmission of $20$ devices and 5\% error probability for each IoT device specific phasing word. In contrary to CKM based cooperative transmission, phasing is now based on side-information and, accordingly, the role of Rice factor is small. Therefore we have only plotted curves for $\nu=-3$dB and $\nu=9$dB. In side-information based approach the volume and quality of side-information is important. As can be seen from Fig. \ref{fig:outage_probability_fb_error}, there is a notable performance difference between the case where only a single phasing bit is send with respect to case where two bits are provided. Yet, increasing further the number of bits in the phasing would not provide much additional performance gain.                 

\begin{figure}
\centering
\includegraphics[width=9cm]{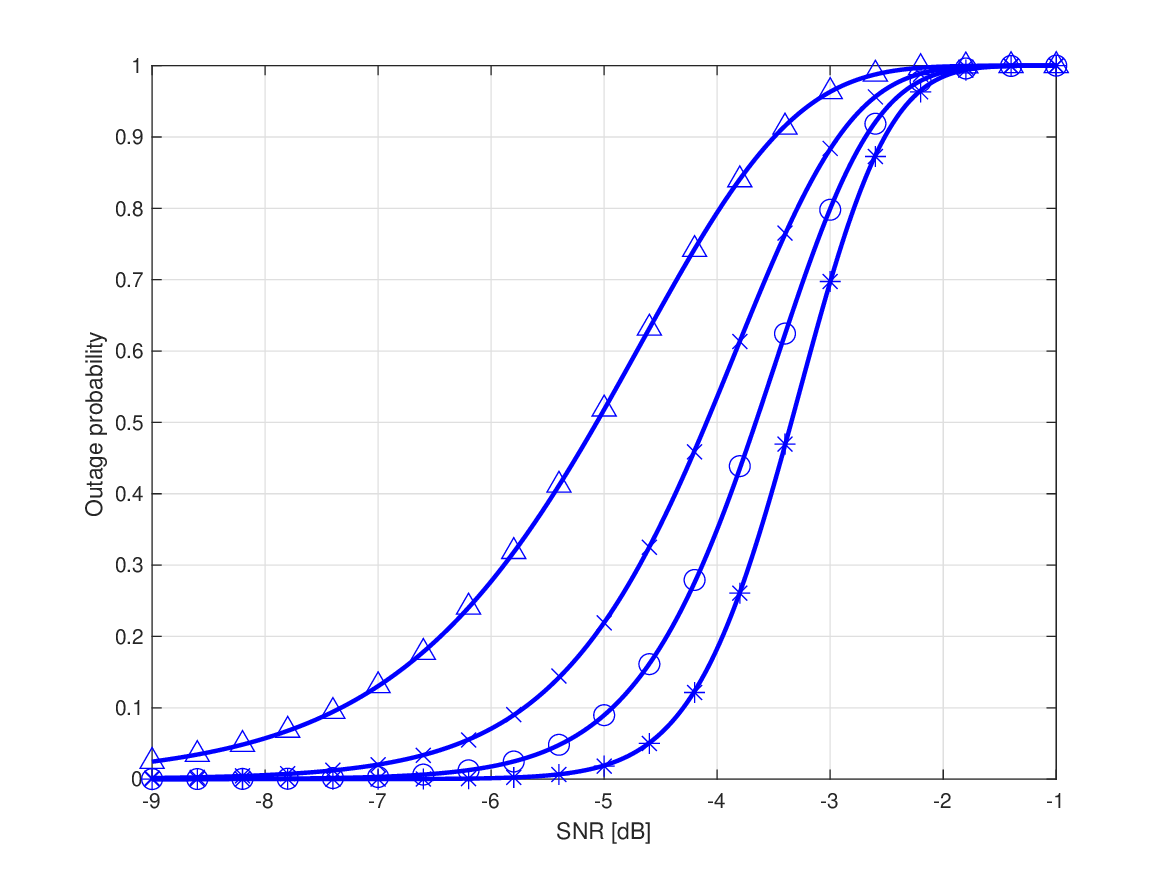}
\caption{Outage probability for side-information based phasing when $|\delta|=20$, $\nu=6$dB and $N=2$. Phasing word error probabilities are $p_w=0.01$ (*), $p_w=0.05$ (o), $p_w=0.10$ (x) and $p_w=0.20$ ($\Delta$). Ticks refer to simulated values and solid curves are plotted using analytic formulae. }
\label{fig:outage_probability_for_different_phasing_word_errors}
\end{figure}

\begin{figure}
\centering
\includegraphics[width=9cm]{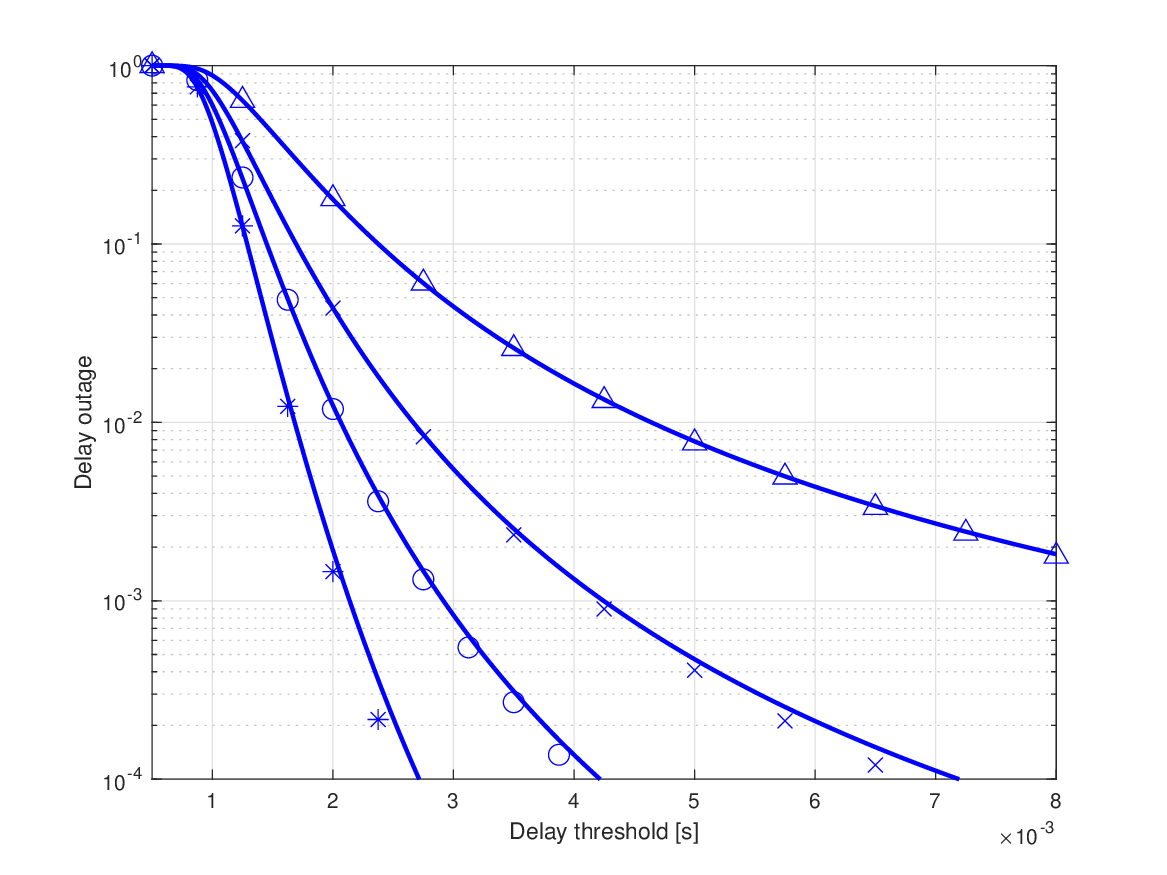}
\caption{Delay outage rate for side-information based phasing when $|\delta|=20$, $\nu=0$dB and $N=2$. Curves: $p_w=0.01$ (*), $p_w=0.05$ (o), $p_w=0.10$ (x), $p_w=0.20$ ($\Delta$). Ticks refer to simulated values and solid curves are plotted using analytic formulae. }
\label{fig:delay_outage_fb_error}
\end{figure}

Fig. \ref{fig:outage_probability_for_different_phasing_word_errors} shows the impact of the impact of word errors in the side-information phasing for each device on the performance when Rice factor is 6dB and $N=2$. Even though erroneous side-information degrades the performance, results indicate that in the range of 10\% or higher probabilities the impact of errors is tolerable. Yet, performance losses increase towards lower probability levels. This can be crucial in case of services that require low latency and high reliability as depicted in Fig. \ref{fig:delay_outage_fb_error}. Therein the impact of errors on the distribution tail is clearly visible. If the DOR level is low, then side-information errors may have a huge impact on the performance in terms of very high delay threshold. Fig. \ref{fig:delay_outage_fb_error} also suggests that incomplete side-information may set a lower bound for the DOR such that certain service requirements cannot be fulfilled at all in case of corrupted side-information. Finally, we note that there are small approximation errors seen in lower DOR levels in Fig. \ref{fig:delay_outage_fb_error}. This is due the fact accurate approximation of the distribution tail becomes increasingly difficult on lower probability levels.       
 
\subsection{Comparison of Scenario 1 and 2 approaches}

Let us briefly compare the approaches used in Scenario 1 and 2. First, we note that CKM based approach has clear advantage since no information exchange between the ICU receiver and individual IoT devices is not required. Basically, IoT devices need to coordinate transmission with cluster head that may provide either the location or the phasing information, depending where the CKM including the mapping from ICU position to phasing is stored. However, the CKM based approach is sensitive to situations where the radio environment includes a scattered/reflected channel part since CKM includes only phasing information related to the static channel component. Thus, the approach of Scenario 1 is feasible only in a radio environment with LoS between IoT devices and ICU. 

The approach used in Scenario 2 is more robust to the radio environment. If timely side-information can be provided reliably to IoT devices, then the performance benefit is notable. For that it is important to notice that ICU should be able to estimate the phase of the channel between IoT device and ICU with some tolerable accuracy. This would require a sounding mechanism or use of pilot signals that will use the IoT device power resources. Also, if received SNR in ICU from an individual IoT device is very low, then even rough phase estimation of the channel between ICU and IoT device might be challenging. It is worth noticing that even when side-information is corrupted, the approach of Scenario 2 works well if delay-tolerant services are considered. However, side-information errors may limit the provision of services with high reliability and low latency.     

\section{Conclusions}

We examined a transmission system where clustered IoT devices collaborated to enhance communication range with the information collecting unit (ICU). Two scenarios were presented: the first utilizing the Channel Knowledge Map (CKM) for phasing information, assuming a channel with a strong static component, and the second based on channel side-information estimated and provided to IoT devices by the ICU. In both scenarios, we considered the presence of errors in the information used for cooperation. Key performance indicators included conventional outage probability and the recently introduced delay outage rate (DOR).

Analytic approximations for SNR distributions in both scenarios, accounting for errors, were derived and verified against simulation results, demonstrating close agreement. The main findings include the significant benefits of cooperation between clustered IoT devices, resulting in increased range to the ICU. However, CKM-based cooperation is sensitive to disruptions in the Line of Sight (LoS) component, diminishing potential gains when LoS is disturbed. The cluster size required for reliable transmission becomes impractically large. On the other hand, channel side-information-based cooperation is robust against changes in the radio environment but susceptible to errors in the channel side-information, potentially limiting its use in scenarios with high-reliability and low-latency requirements.

\section*{Appendix: Derivation of \eqref{eq:number_of_IoTD_scenario1}}
For Marqum Q-function there holds \cite{SiAl2000}:
\begin{equation}\label{eq:marcum_q_estimate}
1-\tfrac12\big(e^{-\frac12(a-b)^2}-e^{-\frac12(a+b)^2}\big)\leq Q_1(a,b),\quad a>b
\end{equation}  
and from the CDF expression in \eqref{eq:CDF_scenario1} we obtain 
\begin{equation}
F_{\hat{\gamma}}(\gamma)\leq \tfrac12\Big(e^{-\frac1{2\sigma_{\tiny{\Sigma}}^2}(\sqrt{\bar{\gamma}_{d,\tiny{\Sigma}}}-\sqrt{\gamma})^2}
-e^{-\frac1{2\sigma_{\tiny{\Sigma}}^2}(\sqrt{\bar{\gamma}_{d,\tiny{\Sigma}}}+\sqrt{\gamma})^2}\Big).
\end{equation}  
Here the latter term in the sum becomes small when either $\nu$ or $\nu|\delta|$ is large since  
\begin{equation}
\tfrac{\bar{\gamma}_{d,\tiny{\Sigma}}}{2\sigma_{\tiny{\Sigma}}^2}=\nu(1+(|\delta|-1)e^{-\sigma_\epsilon^2}).
\end{equation}
That is, if the Rice factor is large or if e.g. $\nu>1$ and the number of cooperative devices is large we can estimate 
\begin{equation}\label{eq:setting_limit}
F_{\hat{\gamma}}(\gamma)\leq \tfrac12 e^{-\frac1{2\sigma_{\tiny{\Sigma}}^2}(\sqrt{\bar{\gamma}_{d,\tiny{\Sigma}}}-\sqrt{\gamma})^2}<P_{\text{dor}},
\end{equation}  
where the latter inequality represents the requirement that the upper bound for CDF is smaller than the target DOR. From \eqref{eq:setting_limit} we obtain  
\begin{equation}\label{eq:solving_delta1}
\sqrt{\bar{\gamma}_{d,\tiny{\Sigma}}}-\sqrt{\gamma}>\sqrt{2\sigma_{\tiny{\Sigma}}^2\log(2P_{\text{dor}})^{-1}}.
\end{equation}  
Following the notation in \eqref{eq:CDF_scenario1} we substitute here $2\sigma_{\tiny{\Sigma}}^2=|\delta|\bar{\gamma}_s$ and approximate
\begin{equation}\label{eq:solving_delta2}
\bar{\gamma}_{d,\tiny{\Sigma}}=\gamma_d|\delta|(1+(|\delta|-1)e^{-\sigma_\epsilon^2})\approx \gamma_d|\delta|^2e^{-\sigma_\epsilon^2}.
\end{equation}
We note that approximation is feasible since $1+(|\delta|-1)e^{-\sigma_\epsilon^2}=|\delta|e^{-\sigma_\epsilon^2}+(1-e^{-\sigma_\epsilon^2})$ and the latter term in the brackets is notably smaller than the first term in cases we focus on. That is, if STD of error is less than 30 degrees and the number of devices is larger than 10. We can now write \eqref{eq:solving_delta1} in the form
\begin{equation}\label{eq:solving_delta3}
\sqrt{\bar{\gamma}_{d}}|\delta|-\sqrt{\bar{\gamma}_s\log(2P_{\text{dor}})^{-1}}\sqrt{|\delta|}-\sqrt{\gamma}>0.
\end{equation}
Formula in the left is a second order polynomial with respect to $\sqrt{|\delta|}$ and, after substituting $\gamma_d=\bar{\gamma}\nu/(1+\nu)$, $\bar{\gamma}_s=\bar{\gamma}/(1+\nu)$ and applying some elementary math, we obtain     
\begin{equation}\label{eq:number_of_IoTD_scenario1_appendix}
|\delta|> -\tfrac{e^{\sigma_{\epsilon}^2}}{4\nu}\log(2P_{\text{dor}})\left(1+\sqrt{1-
4e^{-\frac{\sigma_{\epsilon}^2}{2}}\tfrac{\sqrt{\nu(1+\nu)\gamma/\bar{\gamma}}}{\log(2P_{\text{dor}})}} \right)^2.
\end{equation}
We note that, if transmission power in IoT devices is scaled by number of devices to keep the total transmitted power same regardless the number of cooperating devices, then $\gamma_d=\bar{\gamma}\nu/|\delta|(1+\nu)$ and $\bar{\gamma}_s=\bar{\gamma}/|\delta|(1+\nu)$ in \eqref{eq:solving_delta3} and we obtain requirement
\begin{equation}\label{eq:number_of_IoTD_scenario1_appendix_scaled}
|\delta|> e^{\sigma_{\epsilon}^2}\Big(\sqrt{\tfrac1{\nu}\log(2P_{\text{dor}})^{-1}}+\sqrt{\tfrac{1+\nu}{\nu}\tfrac{\gamma}{\bar{\gamma}}}\Big)^2.
\end{equation}
Finally, we note that requirement 
\begin{equation}\label{eq:condition_for_number_eq_appendix}
T_{th}>D\big(W\log_2(1+\tfrac{\bar{\gamma}\nu}{1+\nu}|\delta|^2e^{-\sigma_{\epsilon}^2})\big)^{-1}
\end{equation} 
is obtained from the condition $a>b$ in \eqref{eq:marcum_q_estimate} by setting $a=\sqrt{\bar{\gamma}_{d,\tiny{\Sigma}}}/\sigma_{\tiny{\Sigma}}$ and $b=\sqrt{\gamma}/\sigma_{\tiny{\Sigma}}$.

\bibliographystyle{IEEE}

\end{document}